\documentclass[letter]{jpsj2} 

\title{Enhanced Superconductivity on the Tetragonal Lattice in FeSe \\ under Hydrostatic Pressure}

\author{Kiyotaka Miyoshi$^1$, Koh Morishita$^1$, Eriko Mutou$^1$, Masatoshi Kondo$^1$, 
Osamu Seida$^1$, Kenji Fujiwara$^1$, Jun Takeuchi$^1$, and Shijo Nishigori$^2$} 
\inst{%
$^1$Department of Material Science, Shimane University, Matsue 690-8504, Japan \\
$^2$Department of Materials Analysis, CIRS, Shimane University, Matsue 690-8504, Japan 
}

\abst{
Superconductivity under pressure in FeSe ($T_{\rm c}$$\sim$7.5 K) has been investigated using single-crystal specimens through the 
measurements of DC magnetization and electrical resistivity. A characteristic three-step increase in $T_{\rm c}$ 
has been found under hydrostatic pressure up to $\sim$34 K above 7 GPa. The structural transition from 
a tetragonal phase to an orthorhombic phase ($T_{\rm s}$$\sim$87 K) is found to disappear at $P$$\sim$2.3 GPa, 
above which $T_{\rm c}$ increases rapidly, suggesting that 
the superconductivity is enhanced by the tetragonal environment. 
Under non-hydrostatic pressure, the increase in $T_{\rm c}$ is suppressed and 
the superconductive volume fraction is considerably reduced above 2 GPa, 
probably owing to the breaking of the tetragonal lattice symmetry by the uniaxial stress. 
The intimate correlation between the enhanced (suppressed) superconductivity and the 
tetragonality (orthorhombicity) in the phase diagram is a common feature of FeSe and other iron-pnictide superconductors.

}

\kword{FeSe, high pressure, superconductivity, DC magnetization}

\begin{document}
\maketitle

The recent discovery of superconductivity in LaFeAsO$_{\rm 1-x}$F$_{\rm x}$ ($T_{\rm c}$=26 K)\cite{kamihara} 
has triggered intensive research to understand the pairing mechanism and 
explore superconductivity in related compounds, leading to a wide variety of 
iron-pnictide superconductors, where 
novel superconductivity is generated in iron-pnictide (Fe-Pn) layers 
consisting of edge-sharing FePn$_4$ tetrahedra. 
Indeed, the geometry of the FePn$_4$ tetrahedron is an important factor for determining $T_{\rm c}$. 
In $R$FeAsO$_{\rm 1-x}$F$_{\rm x}$ ($R$=lanthanoid), it is demonstrated that $T_{\rm c}$ is maximized when the FeAs$_4$ 
tetrahedron forms a regular one, i.e., the As-Fe-As bond angle is $\sim$109.5$^{\circ}$\cite{lee}. 
Another important structural parameter is the pnictogen height $h_{\rm Pn}$ measured from the Fe plane, 
which has been proposed to act as a switch between a high-$T_{\rm c}$ state with nodeless pairing and 
a low-$T_{\rm c}$ state with nodal pairing\cite{kuroki}. 
Recently, it has been found in $R$FeAsO$_{\rm 1-x}$F$_{\rm x}$ that $T_{\rm c}$ is 
a function of not only $h_{\rm As}$ but also the lattice constant, 
showing a characteristic evolution on the $h_{\rm As}$ versus 
lattice constant plane\cite{miyoshi13}. 

The application of pressure is an effective way of varying structural parameters without changing the 
chemical composition. 
In particular, pressure-induced superconductivity is of great interest, because 
we can extract valuable information about 
the structural characteristic that the superconductivity favors. 
It has been found that superconductivity abruptly emerges with the disappearance of the spin-density-wave (SDW) state 
under pressure in AFe$_2$As$_2$ (A=Ba\cite{tori,fukazawa,ishikawa,colo,alireza,yamazaki}, 
Sr\cite{alireza,tori,kotegawa1,kotegawa2,matsubayashi} and Eu\cite{terashima,miclea}), 
while the crystal structure changes from orthorhombic to tetragonal simultaneously with the emergence. 
The emergence of superconductivity accompanied by structural transition
is also observed upon carrier doping in 
AFe$_2$As$_2$\cite{ni} and LaFeAsO\cite{luetkens}. 
The disappearance of superconductivity in the orthorhombic structure can be understood as a manifestation of 
the structure sensitivity of the superconductivity, 
especially to the distortion of the FePn$_4$ tetrahedron from a regular one\cite{lee}. 
In most iron pnictides, superconductivity occurs on the tetragonal lattice. 

Superconductivity in FeSe is an exception, occurring on the orthorhombic structure, which appears below $T_{\rm s}$$\sim$87 K, 
the phase transition temperature from the tetragonal state.\cite{mcqueen,pomjakushina,khasanov,bohmer} 
FeSe has a modest $T_{\rm c}$ of $\sim$7 K\cite{hsu}, and shows no long-range SDW order, unlike AFe$_2$As$_2$ and LaFeAsO, 
however, the coexistence of superconductivity and magnetic order in a short-range scale below $T_{\rm c}$ 
has been found at low pressures\cite{bendele}. 
Although many investigations have been conducted on the pressure dependence of $T_{\rm c}$ by the measurements of 
electrical resistivity and magnetization, the $T_{\rm c}$$-$$P$ relation has not yet been established,   
suggesting that $T_{\rm c}$ increases under pressure and exhibits a maximum ($T_{\rm c}^{\rm max}$)
ranging from 20 to 37 K\cite{masaki,medvedev,garbarino,margadonna,braithwaite,miyoshi09,okabe}. 
The wide variation of $T_{\rm c}^{\rm max}$ is mainly due to the difference in the definition of $T_{\rm c}$. 
It is important to investigate how the superconductivity and the structure change under pressure
in FeSe in order to establish the general trend in the relation between them in iron pnictides 
and gain more insight into the superconducting mechanism. 

In the present work, we have performed DC magnetization and electrical resistivity measurements under pressure 
using single-crystal specimens to clarify the pressure evolution of $T_{\rm c}$ and $T_{\rm s}$ precisely. 
Our DC magnetic measurement under pressure is a reliable technique for revealing thermodynamical 
superconducting critical temperatures and has been successfully 
applied to other systems\cite{miyoshi13,miyoshi09,miyoshi06,miyoshi08}.
In this Letter, we report that $T_{\rm c}$ exhibits a characteristic three-step increase 
up to $\sim$34 K, and that $T_{\rm s}$ decreases with increasing pressure and is extrapolated to zero at $P$$\sim$2.3 GPa, 
above which $T_{\rm c}$ increases rapidly, suggesting that the tetragonal environment enhances the 
superconductivity. The $T$$-$$P$ phase diagrams for 
FeSe and AFe$_2$As$_2$ are found to be analogous to each other. 

Single-crystal specimens of FeSe were grown by a NaCl/KCl flux method similar to that 
described in the literature\cite{zhang}. 
We obtained black and shiny crystals with typical dimensions of 
up to $\sim$1$\times$1$\times$0.05 mm$^3$, as shown in the inset of Fig. 1. 
$T_{\rm c}$ determined from the diamagnetic onset was $\sim$7.5 K at ambient pressure. 
The X-ray diffraction pattern of FeSe single crystals is shown in Fig. 1, 
where ($h$0$l$) peaks are observed, 
indicating that the (101) plane is exposed on the surface of the crystal, 
consistent with earlier studies\cite{zhang,hu}. 
For the magnetic measurements under high pressure, a miniature diamond anvil cell (DAC) with an outer 
diameter of 8 mm was used to generate high pressure, which was 
combined with a sample rod of a commercial SQUID magnetometer. 
Details of the DAC are given elsewhere.\cite{mito}
A single crystal of FeSe was loaded into the gasket hole together 
with a small piece of high-purity lead (Pb) [the inset of Fig. 2(d)] 
to realize the $in$ $situ$ observation of pressure
by determining the pressure from the $T_{\rm c}$ shift of Pb. 
Magnetization data for the small amounts of FeSe and Pb were obtained by subtracting the 
magnetic contribution of DAC measured in an empty run from the total magnetization. 
To establish the intrinsic $T_{\rm c}$$-$$P$ relation under hydrostatic pressure up to 8 GPa, 
the measurements were mainly carried out using liquid Ar as the pressure-transmitting medium (PTM). 
To investigate the effect of uniaxial stress, we also used NaCl powder as the PTM. 
Electrical resistivity was measured by a four-probe technique using a clamp-type 
piston-cylinder pressure cell with Daphne oil 7373 as the PTM. 
\begin{figure}[t]
\begin{center}
\includegraphics[width=7.5cm]{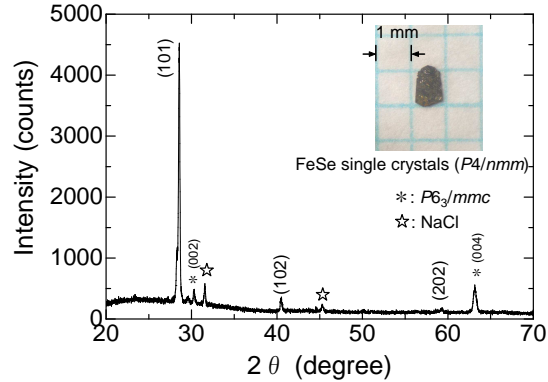}
\end{center}
\caption{(Color on line) Powder X-ray diffraction pattern of FeSe single crystals at room temperature. 
Diffraction peaks corresponding to a tetragonal crystal structure with the symmetry group $P4/nmm$ are 
observed. Those corresponding to a hexagonal structure with $P6_{3}mmc$ are also observed in addition to the 
peaks of NaCl. The inset shows a typical FeSe single crystal.  
}
\label{f1}
\end{figure}

We show the temperature dependence of zero-field-cooled DC magnetization ($M$) under various pressures up to 1.9 GPa using 
Daphne oil 7373 as the PTM in Fig. 2(a) and up to higher pressure using liquid Ar as the PTM in Figs. 2(b) and 2(c). 
In Fig. 2(a), the $M$$-$$T$ curve shows a rapid decrease below 9.5 K at $P$=0.39 GPa. 
The onset of the diamagnetic response ($T_{\rm c}^{\rm diamag}$), 
which is estimated by extrapolating the initial slope of the $M$$-$$T$ curve 
just below the decrease to the normal state magnetization, is defined as $T_{\rm c}$.
In the figure, $T_{\rm c}$ shifts to a higher value at $P$=0.85 GPa but decreases 
to $\sim$10 K at 1.6 GPa and increases again to $\sim$13 K at 1.9 GPa. 
It is shown that $T_{\rm c}$ increases rapidly above $\sim$2 GPa in Fig. 2(b) 
and saturates at $\sim$25 K above $\sim$3 GPa, but gradually increases again 
above $\sim$5 GPa to $T_{\rm cmax}$$\sim$34 K in Fig. 2(c). 
We show plots of $T_{\rm c}$ versus $P$ in Fig. 2(d), 
where the $T_{\rm c}$$-$$P$ curve exhibits a characteristic three-step increase. 
It should be noted that the second increase in $T_{\rm c}$ above $\sim$2 GPa 
is extremely rapid, with $dT_{\rm c}$$/$$dP$$\sim$ 10 K/GPa, 
compared with that seen in all other superconductors, implying the occurrence of a pressure-induced 
phase transition. The inset of Fig. 2(d) shows the $T_{\rm c}$$-$$P$ curves for FeSe obtained 
through the resistivity measurements under hydrostatic pressure using a cubic anvil press by Okabe $et$ $al$.,\cite{okabe} 
showing that the onset temperature of the 
resistivity drop ($T_{\rm c}^{\rm onset}$) exhibits a maximum of $\sim$38 K, considerably larger than 
that of the zero-resistive temperature ($T_{\rm c}^{\rm zero}$) of $\sim$30 K.   
The behavior of the $T_{\rm c}$$-$$P$ curve observed in the present study is quantitatively 
consistent with that of the $T_{\rm c}^{\rm zero}$$-$$P$ curve 
below 5 GPa but clearly different from that above 6 GPa. 
In contrast, the behavior is fundamentally different from that of the $T_{\rm c}^{\rm onset}$$-$$P$ curve. 
Since the sharpness of the transition is almost unchanged even at 7.5 GPa as shown in Fig. 2(c), 
the $T_{\rm c}$$-$$P$ curve in Fig. 2(d) is thought to be intrinsic, 
while $T_{\rm c}$ may be overestimated by defining $T_{\rm c}^{\rm onset}$ as $T_{\rm c}$, 
as in previous reports\cite{medvedev,margadonna}. 
\begin{figure}[t]
\begin{center}
\includegraphics[width=7.5cm]{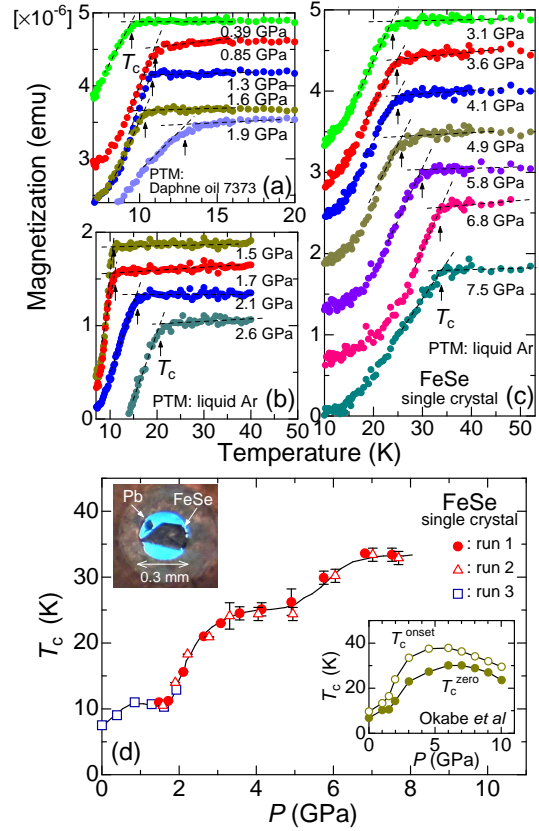}
\end{center}
\caption{(Color on line) 
Temperature dependence of zero-field-cooled DC magnetization for FeSe single crystal measured in 
a magnetic field of 20 Oe under various pressures 
using Daphne oil 7373 (a) and liquid Ar (b), (c) 
as the pressure transmitting medium (PTM). The data are intentionally shifted along the longitudinal axis for clarity. 
(d) $T_{\rm c}$$-$$P$ curve for FeSe single crystals. The inset shows the $T_{\rm c}$$-$$P$ curves 
reported by Okabe $et$ $al$\cite{okabe}. 
obtained through the measurements of electrical resistivity under pressure. A photo of a FeSe single crystal and 
a small piece of Pb set in the Cu-Be gasket hole is also shown.  
}
\label{f1}
\end{figure}

Next, we show the temperature dependence of electrical resistivity ($\rho$) under pressure in Fig. 3(a). 
Metallic behavior with a dip at $\sim$87 K and a sharp superconducting transition are 
observed. The dip anomaly corresponds to the structural phase 
transition from a tetragonal state to an orthorhombic state at $T_{\rm s}$$\sim$87 K. 
It is shown that $T_{\rm s}$ decreases with increasing pressure down to $\sim$40 K at 1.8 GPa, 
indicating that the transition vanishes with increasing pressure. 
In Fig. 3(b), $T_{\rm c}^{\rm zero}$ is found to increase nonmonotonically with pressure. 
In AFe$_2$As$_2$, a relatively sharp anomaly compared with 
that seen in Fig. 3(a) is observed at $T_{\rm s}$ in the 
$\rho$$-$$T$ curve (see, e.g., Ref. 13), since the effect of the magnetic scattering on the resistivity is 
abruptly reduced owing to the antiferromagnetic ordering below $T_{\rm s}$. 
To determine the position of $T_{\rm s}$, the $\rho$($T$) data after 
subtracting the linear component are plotted in Fig. 3(c). 
We defined the bottom of the dip as $T_{\rm s}$. 
In Fig. 3(d), the $T_{\rm s}$ versus pressure data are plotted together with the $T_{\rm c}^{\rm zero}$($P$) 
data and $T_{\rm c}^{\rm diamag}$($P$) curve to construct the $T$$-$$P$ phase diagram for FeSe. 
As seen in the figure, the orthorhombic phase is expected to disappear above $P$$\sim$2.3 GPa, 
where superconductivity is changed to be developed on the tetragonal lattice, 
accompanied by a large increase in $T_{\rm c}$ of $dT_{\rm c}$$/$$dP$$\sim$ 10 K/GPa. 
The inset of Fig. 3(d) shows the $T$$-$$P$ phase diagram 
for SrFe$_2$As$_2$ reported by Matsubayashi $et$ $al$\cite{matsubayashi}. 
A common feature of the phase diagrams is the intimate correlation between the enhanced (suppressed) 
superconductivity and the tetragonal (orthorhombic) lattice structure, 
which is also seen in AFe$_2$As$_2$\cite{ni} and LaFeAsO\cite{luetkens} upon carrier doping. 
Thus, it is a general trend in iron-pnictide superconductors 
that the superconductivity favors a tetragonal environment but disfavors an orthorhombic one. 
In FeSe, superconductivity survives even in the orthorhombic structure, in contrast to that in AFe$_2$As$_2$ and LaFeAsO.
This is due to the absence of long-range magnetic order in the orthorhombic structure, 
while the SDW state appears in AFe$_2$As$_2$ and LaFeAsO, competing with superconductivity. 
The strong enhancement of the superconductivity in FeSe above $\sim$2 GPa therefore appears to be 
of purely structural origin, and is not due to the disappearance of a competing magnetic ground state. 
However, we should note the coexistence of static magnetic order and superconductivity, 
which has been observed above 0.8 GPa in the $\mu$SR experiments\cite{bendele}, although it is unclear 
whether or not the magnetic order vanishes above 2 GPa. 

\begin{figure}[t]
\begin{center}
\includegraphics[width=7.5cm]{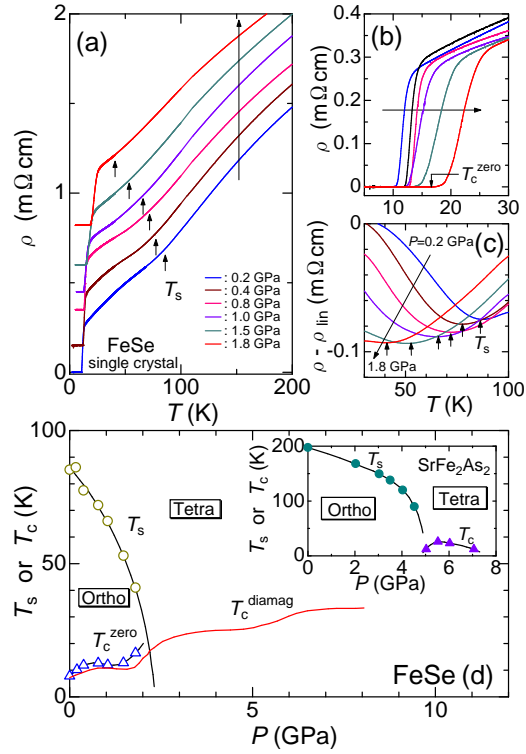}
\end{center}
\caption{(Color on line) 
Temperature dependence of electrical resistivity $\rho$ for FeSe under various pressures (a) 
and an enlarged view below 30 K (b). (c) Data after subtracting the linear component $\rho_{\rm lin}$, 
showing a dip anomaly under various pressures. The bottom temperature is defined 
as $T_{\rm s}$, the transition temperature 
from a tetragonal phase to an orthorhombic phase. (d) Temperature-pressure phase diagram for FeSe. 
The inset shows that for SrFe$_2$As$_2$ obtained by Matsubayashi $et$ $al$\cite{matsubayashi}. $T_{\rm s}$ and  
$T_{\rm c}$ determined from the onset of the diamagnetic response ($T_{\rm c}^{\rm diamag}$) 
and the zero-resistivity temperature ($T_{\rm c}^{\rm zero}$) are shown as a function of pressure.  
}
\label{f1}
\end{figure}
 
In our previous study, we observed an abrupt reduction in the amplitude of diamagnetic response 
in FeSe above 2 GPa using polycrystalline specimens and Daphne oil 7373 as the PTM, 
and encountered difficulty in detecting the superconducting transition at higher pressure\cite{miyoshi09}.  
This behavior can be understood as a result of superconductivity being suppressed in some parts of 
the specimen by the application of uniaxial pressure, which is realized by the solidification of 
Daphne oil 7373 at high pressures above 2.2 GPa, while liquid Ar generates high pressure with good 
hydrostaticity up to 10 GPa\cite{tateiwa}. Uniaxial stress anisotropically compresses the crystal, 
possibly breaking the tetragonal crystal symmetry, and can lead to the suppression of superconductivity. 
\begin{figure}[t]
\begin{center}
\includegraphics[width=7.5cm]{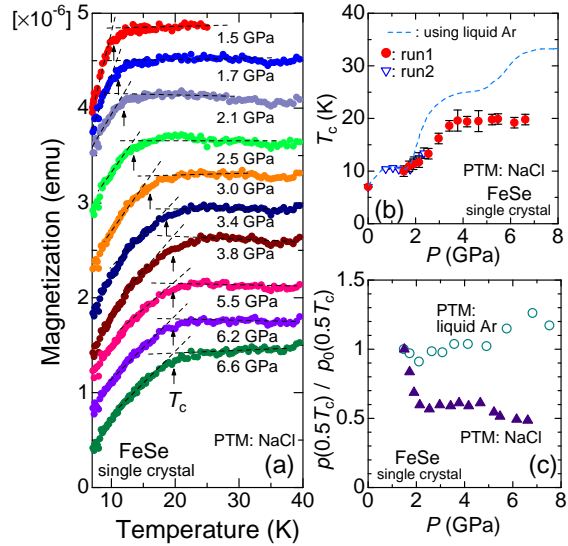}
\end{center}
\caption{(Color on line) 
Temperature dependence of zero-field-cooled DC magnetization under various pressures measured with 
a magnetic field of 20 Oe 
(a) and plots of $T_{\rm c}$ versus pressure (b) for FeSe single crystals using NaCl as the pressure transmitting medium (PTM). 
The broken curve in (b) is the $T_{\rm c}$$-$$P$ curve observed by the measurements using liquid Ar as the PTM. 
(c) Plots of superconducting volume fraction ($p$) at $T$$=$0.5$T_{\rm c}$ normalized to that at $P$=1.5 GPa ($p_{0}$) 
versus pressure. 
}
\label{f1}
\end{figure}

To confirm the effect of uniaxial stress on the superconductivity in FeSe, magnetic 
measurements were performed on the single crystals using NaCl powder as the solid-state PTM. 
In the experiments, the surface of a single crystal was set parallel to a diamond culet so that 
uniaxial stress could be applied on the (101) plane to compress the crystal along both the $a$- and $c$-axes.  
We show the $M$$-$$T$ data measured using NaCl under various pressures in Fig. 4(a), 
where $T_{\rm c}$ increases with increasing pressure up to 20 K at $P$=3.8 GPa and then 
becomes pressure-independent up to $P$=6.6 GPa. The $T_{\rm c}$$-$$P$ relation is shown 
in Fig. 4(b), together with the $T_{\rm c}$$-$$P$ curve under hydrostatic pressure using Ar as the PTM. 
Also, the superconducting volume fraction estimated from the amplitude of the diamagnetic response, 
$p$, at $T$=0.5$T_{\rm c}$ normalized to that at 1.5 GPa ($p_{0}$) is plotted in Fig. 4(c). 
We note that the difference between the $T_{\rm c}$ values under hydrostatic and uniaxial pressures 
is large above $\sim$2 GPa, and even larger above $\sim$5 GPa. 
Furthermore, the volume fraction $p$ is roughly constant under hydrostatic pressure but 
rapidly decreases by half under uniaxial pressure, as similarly seen in our previous study\cite{miyoshi09}. 
The small increase in $T_{\rm c}$ and the reduction in the volume fraction observed above 2 GPa are 
attributable to the breaking of the tetragonal lattice symmetry due to the uniaxial stress along the (101) direction.  
The volume fraction is expected to be even lower for the compression along the (100) direction, 
considering that the volume fraction for the polycrystalline specimens is reduced to $\sim$1/5 above 2$-$3 GPa\cite{miyoshi09}.  

In summary, we have investigated the superconductivity and the structural phase transition in FeSe under pressure 
by the measurements of dc magnetization and electrical resistivity using single-crystal specimens. 
It has been found that $T_{\rm c}$ increases in three steps with the application of hydrostatic pressure up to $\sim$34 K and 
the orthorhombic phase disappears above $P$$\sim$2.3 GPa. 
In FeSe, the superconductivity on the orthorhombic lattice changes to occur on the tetragonal lattice 
under hydrostatic pressure, accompanied by an extremely rapid increase in $T_{\rm c}$. 
Under uniaxial pressure, the increase in $T_{\rm c}$ is suppressed to $\sim$20 K and 
the volume fraction of the superconductivity is considerablty reduced above 2 GPa, 
probably owing to the breaking of the tetragonal symmetry. 
The intimate correlation between the enhanced (suppressed) superconductivity and the tetragonal (orthorhombic) 
structure in the phase diagram is a common feature of iron-pnictide superconductors.



\begin{acknowledgment}
The authors thank Mr. H. Katsube for his technical assistance. 

\end{acknowledgment}


\begin{thebibliography}{9}
\bibitem{kamihara} Y. Kamihara, T. Watanabe, M. Hirano, and H. Hosono, J. Am. Chem. Soc. \textbf{130}, 3296 (2008).

\bibitem{lee} C.-H. Lee, A. Iyo, H. Eisaki, H. Kito, M. T. Fernandez-Diaz, T. Ito,
K. Kihou, H. Matsuhata, M. Braden, and K. Yamada, J. Phys. Soc. Jpn. {\bf 77}, 083704 (2008). 

\bibitem{kuroki} K. Kuroki, H. Usui, S. Onari, R. Arita, and H. Aoki, Phys. Rev. B {\bf 79}, 224511 (2009). 

\bibitem{miyoshi13} K. Miyoshi, E. Kojima, S. Ogawa, Y. Shimojo, and J. Takeuchi, Phys. Rev. B {\bf 87}, 235111 (2013). 

\bibitem{tori} M. S. Torikachvili, S. L. Bud'ko, N. Ni, and P. C. Canfield, Phys. Rev. B {\bf 78}, 104527 (2008).

\bibitem{fukazawa} H. Fukazawa, N. Takeshita, T. Yamazaki, K. Kondo, K. Hirayama, Y. Kohori, K. Miyazawa, 
H. Kito, H. Eisaki, and A. Iyo, J. Phys. Soc. Jpn. {\bf 77}, 105004 (2008).

\bibitem{ishikawa} F. Ishikawa, N. Eguchi, M. Kodama, K. Fujimaki, M. Einaga, A.
Ohmura, A. Nakayama, A. Mitsuda, and Y. Yamada, Phys. Rev. B {\bf 79}, 172506 (2009).

\bibitem{colo} E. Colombier, S. L. Bud'ko, N. Ni, and P. C. Canfield, Phys. Rev. B {\bf 79}, 224518 (2009).

\bibitem{alireza} P. L. Alireza, Y. T. C. Ko, J. Gillett, C. M. Petrone, J. M. Cole, G. G. Lonzarich, and S. E. Sebastian, 
J. Phys.: Condens. Matter {\bf 21}, 012208 (2009). 

\bibitem{yamazaki} T. Yamazaki, N. Takeshita, R. Kobayashi, H. Fukazawa, Y. Kohori, K. Kihou, C.-H. Lee, 
H. Kito, A. Iyo, and H. Eisaki, Phys. Rev. B {\bf 81}, 224511 (2010). 

\bibitem{kotegawa1} H. Kotegawa, H. Sugawara, and H. Tou, J. Phys. Soc. Jpn. {\bf 78}, 013709 (2009). 

\bibitem{kotegawa2} H. Kotegawa, T. Kawazoe, H. Sugawara, K. Murata, and H. Tou, 
J. Phys. Soc. Jpn. {\bf 78}, 083702 (2009).

\bibitem{matsubayashi} K. Matsubayashi, N. Katayama, K. Ohgushi, A. Yamada, K. Munakata,
T. Matsumoto, and Y. Uwatoko, J. Phys. Soc. Jpn. {\bf 78}, 073706 (2009).

\bibitem{terashima} T. Terashima, M. Kimata, H. Satsukawa, A. Harada, K. Hazama, S. Uji, 
H. S. Suzuki, T. Matsumoto, and K. Murata, J. Phys. Soc. Jpn. {\bf 78}, 083701 (2009). 

\bibitem{miclea} C. F. Miclea, M. Nicklas, H. S. Jeevan, D. Kasinathan, Z. Hossain,
H. Rosner, P. Gegenwart, C. Geibel, and F. Steglich, Phys. Rev. B {\bf 79}, 212509 (2009).

\bibitem{ni} N. Ni, M. E. Tillman, J.-Q. Yan, A. Kracher, S. T. Hannahs, S. L. Bud'ko, and P. C. Canfield, 
Phys. Rev. B {\bf 78}, 214515 (2008).

\bibitem{luetkens} H. Luetkens, H.-H. Klauss, M. Kraken, F. J. Litterst, T. Dellmann, R. Klingeler, C. Hess, 
R. Khasanov, A. Amato, C. Baines, M. Kosmala, O. J. Schumann, M. Braden, J. Hamann-Borrero, N. Leps, A. Kondrat, 
G. Behr, J. Werner, and B. B$\ddot{\rm u}$chner, Nat. Mater. {\bf 8}, 305 (2009).

\bibitem{mcqueen} T. M. McQueen, A. J. Williams, P. W. Stephens, J. Tao, Y. Zhu, V. Ksenofontov, F. Casper, 
C. Felser, and R. J. Cava, Phys. Rev. Lett. {\bf 103}, 057002 (2009).

\bibitem{pomjakushina} E. Pomjakushina, K. Conder, V. Pomjakushin, M. Bendele, and R. Khasanov, 
Phys. Rev. B {\bf 80}, 024517 (2009). 

\bibitem{khasanov}  R. Khasanov, M. Bendele, K. Conder, H. Keller, E. Pomjakushina,
and V. Pomjakushin, New J. Phys. {\bf 12}, 073024 (2010). 

\bibitem{bohmer} A. E. B$\ddot{\rm o}$hmer, F. Hardy, F. Eilers, D. Ernst, P. Adelmann, P. Schweiss, T. Wolf, and C. Meingast, 
Phys. Rev. B {\bf 87}, 180505(R) (2013). 

\bibitem{hsu} F.-C. Hsu, J.-Y. Luo, K.-W. Yeh, T.-K. Chen, T.-W. Huang, P. M. Wu, Y.-C. Lee, Y.-L. Hung, 
Y.-Y. Chu, D.-C. Yan, and M.-K. Wu, Proc. Natl. Acad. Sci. U.S.A. {\bf 105}, 14262 (2008).

\bibitem{bendele}  M. Bendele, A. Amato, K. Conder, M. Elender, H. Keller, H.-H. Klauss, H. Luetkens, E. Pomjakushina, 
A. Raselli, and R. Khasanov, Phys. Rev. Lett. {\bf 104}, 087003 (2010). 

\bibitem{masaki} S. Masaki, H. Kotegawa, Y. Hara, H. Tou, K. Murata, Y. Mizuguchi,
and Y. Takano, J. Phys. Soc. Jpn. {\bf 78}, 063704 (2009). 

\bibitem{medvedev} S. Medvedev, T. M. McQueen, I. A. Troyan, T. Palasyuk, M. I.
Eremets, R. J. Cava, S. Naghavi, F. Casper, V. Ksenofontov, G.
Wortmann, and C. Felser, Nat. Mater. {\bf 8}, 630 (2009).

\bibitem{garbarino} G. Garbarino, A. Sow, P. Lejay, A. Sulpice, P. Toulemonde, W. Crichton, M. Mezouar, 
and M. N$\acute{\rm a}$$\tilde{\rm n}$ez-Regueiro, Europhys. Lett. \textbf{86}, 27001 (2009). 

\bibitem{margadonna} S. Margadonna, Y. Takabayashi, Y. Ohishi, Y. Mizuguchi, Y.
Takano, T. Kagayama, T. Nakagawa, M. Takata, and K. Prassides,
Phys. Rev. B {\bf 80}, 064506 (2009).

\bibitem{braithwaite} D. Braithwaite, B. Salce, G. Lapertot, F. Bourdarot, C. Marin, D. Aoki, 
and M. Hanfland, J. Phys.: Condens. Matter {\bf 21}, 232202 (2009).

\bibitem{miyoshi09} K. Miyoshi, Y. Takaichi, E. Mutou, K. Fujiwara, and J. Takeuchi,
J. Phys. Soc. Jpn. {\bf 78}, 093703 (2009).

\bibitem{okabe} H. Okabe, N. Takeshita, K. Horigane, T. Muranaka, and J. Akimitsu, 
Phys. Rev. B {\bf 81}, 205119 (2010).

\bibitem{miyoshi06} K. Miyoshi, Y. Takamatsu, and J. Takeuchi, 
J. Phys. Soc. Jpn. \textbf{75}, 065001 (2006). 

\bibitem{miyoshi08} K. Miyoshi, Y. Takaichi, Y. Takamatsu, M. Miura, and J. Takeuchi, 
J. Phys. Soc. Jpn. \textbf{77}, 043704 (2008). 

\bibitem{zhang} S. B. Zhang, X. D. Zhu, H. C. Lei, G. Li, B. S. Wang, L. J. Li,
X. B. Zhu, Z. R. Yang, W. H. Song, J. M. Dai, and Y. P. Sun, Supercond. Sci. Technol. {\bf 22}, 075016 (2009). 

\bibitem{hu} R. Hu, H. Lei, M. Abeykoon, E. S. Bozin, S. J. L. Billinge, 
J. B. Warren, T. Siegrist, and C. Petrovic, Phys. Rev. B {\bf 83}, 224502 (2011).

\bibitem{mito} M. Mito, M. Hitaka, T. Kawae, K. Takeda, T. Kitai, and N. Toyoshima, 
Jpn. J. Appl. Phys. {\bf 40}, 6641 (2001). 

\bibitem{tateiwa} N. Tateiwa and Y. Haga, Rev. Sci. Instrum. {\bf 80}, 123901 (2009). 

\end{thebibliography}
\end{document}